# Adaptive sampling of pressure distribution features for the airfoil statistical analysis

Runze Li,[*] Yufei Zhang,[†] Haixin Chen[‡]

*Tsinghua University, Beijing, 100084, People's Republic of China*

**In the area of supercritical wing design, a variety of principles, laws and rules have been summarized by scholars who perform theoretical and experimental analyses. The applicability of these rules is usually restricted by the airfoil samples investigated. With the advance of computational fluid dynamics and computational intelligence, such work can be better conducted on computers. The present paper proposes an output space sampling method to generate airfoil samples that have specified pressure distributions or meet certain special requirements. The well-selected and distributed samples are then utilized for statistical studies to obtain more reliable or more universal aerodynamics rules that can be used as guidance in the process of supercritical airfoil design. The capabilities of the output space sampling method in regard to filling the space and exploring the boundaries are also improved. The output space sampling method is employed to generate supercritical airfoil samples with different requirements, allowing the relationships between the shock wave location and the drag divergence Mach number as well as the drag creep characteristic to be revealed.**

## Nomenclature

*AoA*     = Angle of attack

*Re*     = Reynolds number

*c*     = Airfoil chord length

$C_L$     = Lift coefficient

---

[*] Ph. D. student, School of Aerospace Engineering, email: lirz16@mails.tsinghua.edu.cn
[†] Associate professor, School of Aerospace Engineering, senior member AIAA, email: zhangyufei@tsinghua.edu.cn
[‡] Professor, School of Aerospace Engineering, associate fellow AIAA, email: chenhaixin@tsinghua.edu.cn

| | |
|---|---|
| $C_d$ | = Drag coefficient |
| $C_p$ | = Pressure coefficient |
| $d$ | = Distance |
| $Err$ | = Smoothness function of the suction plateau |
| $k$ | = Critical slope of $C_d - M_\infty$ curve for drag divergence |
| $M_w$ | = Wall Mach number |
| $M_\infty$ | = Free-stream Mach number |
| $M_{DD}$ | = Drag divergence Mach number |
| **x** | = Independent variables |
| $X$ | = Location |
| **y** | = Dependent variables, i.e., outputs |
| $t$ | = Airfoil thickness |

## I. Introduction

The supercritical airfoil offers superior performance for transonic flight, including a larger cruise L/D, a delayed drag divergence Mach number, and a large volume for the structure and fuel storage. These benefits can be attributed to the airfoil's performance in terms of establishing an appropriate pressure distribution, which balances the shock wave's drag penalty and its benefits toward pressure recovery, striking a compromise between lift generation and the Mach number in front of the shock wave. The complexity of the pressure distribution lies in the nonlinear and sensitive nature of transonic aerodynamics, which is caused by the on-body shock wave and its interaction with the boundary layer.

Over the past 60 years, many researchers have both theoretically and experimentally studied the relationships between airfoil performance and these flow features. Their work has significantly contributed to aircraft design [1-3]. Korn's equation [4] describes the growth of the shock wave relative to the free-stream Mach number and further states the relationship among the thickness, drag divergence Mach number, and lift coefficient. Oswatitsch's theorem shows that the wall Mach number in front of a shock wave can be used to estimate wave drag [5] and can be applied in optimization design [6]. Zhang studied the effects of the pressure gradient of a suction plateau on supercritical natural laminar flow airfoils [7]. Pearcey [8] investigated the relationship between the shock wave and incipient separation, which occurs when a reverse flow is imminent in the boundary layer. Furthermore, the size of a shock wave-induced boundary layer separation was also studied to predict the buffet onset lift coefficient [9], and the spanwise location of the separation was found to contribute to the buffet severity [10]. In combination, these studies revealed the impact of flow features on performance and showed the importance of these relationships to aircraft design.

These relations were mostly concluded based on theoretical or experimental studies of several typical conditions and samples. With the development of computational fluid dynamics (CFD), CFD-based statistical research has started to gain additional attention. Emerging technologies such as machine learning and data mining can help researchers to effectively obtain desired samples for subsequent analysis. Currently, researchers have exploited CFD data via surrogate models in aerodynamic optimizations [11-13]. These methods, which include surrogate-based optimization methods [14] and data-driven approaches [15], utilize data to predict performances based on the geometry of a wing or an airfoil. Even so, the relationships between aerodynamic performance and geometry are typically too complicated, or incomprehensible, for effective analysis. In contrast, although the relationships between performance and flow features are more intuitive and useful for aircraft designs, most studies are still based on the results of several typical cases of experiments or numerical simulations, which could make the conclusions incomplete or even biased. The main reason for the lack of comprehensive and unbiased statistical studies is that it is difficult to provide sufficient, typical and well-distributed samples.

For the statistical studies of flow features, sufficient samples with different flow features must be provided to the research community. That is, instead of the geometry space, it is the "feature space" that must be sampled in. The geometry coefficients are usually represented by the input vector **x**, since flow features are governed by Navier-Stokes equations and must be solved in terms of specific geometries. Therefore, the flow features and performances are the output vectors, represented by **y** and **Y**, respectively, as shown in Figure 1. Many sampling methods have been developed either to achieve a good space-filling property of the input space or to improve the quality of the surrogate models to approximate the high-dimensional nonlinear relationship **x** → **y** (or **Y**) [17]. However, because the outputs cannot be directly assigned by sampling methods when studying the **y** → **Y** relationships, a novel Output Space Sampling (OSS) method is needed to generate samples with various flow features (**y**). This is a new challenge for which the conventional sampling methods have not been designed.

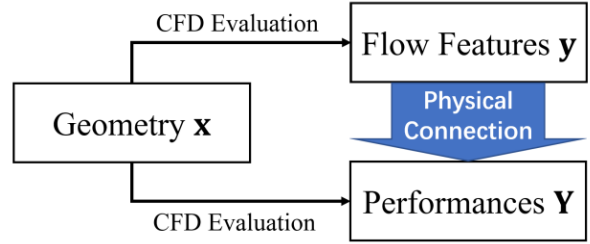

**Figure 1 Geometry, flow features, and performances**

In this paper, statistical studies of supercritical airfoils are carried out to prove that the proposed OSS method can generate airfoil samples with different flow features. For supercritical airfoils, pressure distributions can imply most of the flow information and mechanisms, of which the flat-top suction plateau, weak shock, and aft-loading, are the most important features. For instance, the location and strength of the shock wave, the slope of pressure recovery and the extent of aft-loading can nearly be used to determine the lift to drag ratio (L/D), drag divergence, buffeting and stall properties of an airfoil. During aircraft design processes, engineers observe and manipulate pressure distributions based on their knowledge of the relationship between pressure distributions and the aircraft performances. Such knowledge is sometimes obtained from aerodynamics textbooks; however, due to the complexity of transonic viscous flows, most of this knowledge must be obtained from experience. In either case, it would be very beneficial if the relationships could be universally correct and unbiased. Then, even though statistical studies are regarded as an advanced means of accumulating experience and knowledge, it places strong demands on samples. To ensure accuracy, the samples used for statistical studies must cover different types of pressure distributions and various combinations of pressure distribution features.

There are several methods for obtaining the desired pressure distribution of feature samples, e.g., inverse design methods [16] and pressure distribution guided optimizations [6]. Both types of approaches can determine the geometry when certain flow features are specified. However, an optimization searching process is required for each desired pressure distribution, which can lead to significant computational costs. Therefore, an effective output space sampling method is still desired.

This paper proposes an OSS method that utilizes an adaptive sampling algorithm to generate samples with different flow features. Then, by generating airfoil samples with well-distributed pressure distribution features, the correlations between the shock wave and several airfoil performances are studied statistically.

The paper is organized as follows. First, the relevant background is discussed. Second, the output space sampling algorithm is described, and several sampling criteria are proposed for either improving the output space-filling property or exploring the output space boundaries. Third, the OSS method is then compared with other sampling methods by using analytical test functions. Fourth, the OSS method is used to generate airfoils with various pressure distributions, and these samples are then statistically studied. Finally, the results are discussed and found to indicate that the drag divergence Mach number increases when the airfoil has a more upstream shock wave, whereas the drag creep characteristic may deteriorate if the shock wave occurs too far upstream.

## II. Output space sampling method

Sampling methods are essential tools for statistical studies and data mining. Most sampling methods are developed for exploration and exploitation, i.e., exploring the most unexplored region and refining the search in the regions near existing samples. The conventional sampling methods usually aim either to fill the input space with the greatest efficiency when using uniformly distributed samples or to arrange samples according to a specified probability density distribution, so that the samples are valid for statistical studies. In other cases, the conventional sampling methods are designed to improve surrogate models of $\mathbf{x} \to \mathbf{y}$ (or $\mathbf{Y}$); in such instances, the methods tend to add samples to unexplored places or places with rapid changes in the function value. However, when conducting statistical studies regarding the relationships among $\mathbf{y}$ and $\mathbf{Y}$, the sample distribution in the output space is the focus. Although the conventional methods can manage exploration and exploitation, nonetheless the generated samples do not necessarily have the desired distribution in the output space. Therefore, novel methods to conduct output space samplings are required.

In this section, an OSS method is proposed. The method is developed from a basic adaptive sampling algorithm by utilizing differential evolution optimizations and several novel criteria. The OSS method is designed to tackle three main challenges in the statistical studies of outputs. The first challenge is to indirectly sample the outputs by finding the corresponding $\mathbf{x}$. The second challenge is to generate samples according to a specified sample distribution when the constraints of the samples are given. The third challenge concerns the output space boundaries: sometimes not only does the output space-filling property require improvement, but the output space boundaries must also be explored because they cannot be known beforehand.

This section begins with an introduction to conventional input space sampling methods. Then, the adaptive sampling algorithm used for OSS is described. The conventional criteria and Euclidean distance criteria for the OSS method are introduced. Because there are several drawbacks associated with these criteria, a Target Shooting Criterion (TSC) is proposed to effectively and efficiently manipulate the sample distribution in the output space. Then, the OSS method is tested and compared with other methods by using an analytical test function.

**A. Background of conventional sampling methods**

Sampling methods can be roughly categorized into stationary and adaptive sampling methods [17]. Stationary sampling methods solely focus on the input space and tend to generate samples that are uniformly spread in the input space. The Latin Hypercube Sampling (LHS) method [18] is a commonly used stationary sampling method, but its space-filling behavior is not always guaranteed. Several modifications have been developed to improve its performance, and the improved Latin Hypercube Sampling (iLHS) method can achieve sufficient input space-filling properties for most situations [19]. Euclidean distance-based designs [20, 21] use criteria that only rely on input space information, such as maximin distance and crowding distance metrics, to generate samples. The Audze-Eglais design used the "potential energy" criterion to represent the space-filling property in the input space [22]. These criteria can gain a good space-filling property by minimizing the "potential energy" or maximizing the minimum distance between samples, although the cost of optimizing the sample distribution significantly increases with the growing size of the sample set. Therefore, these criteria are usually used to add points in adaptive methods for design space exploration.

Adaptive sampling methods are usually based on surrogate models and thus are usually conducted to strengthen the surrogate model or increase the space-filling property. The Kriging model and radial basis function (RBF) response surface are the two most commonly used surrogate models. Adaptive sampling methods typically begin with a relatively small number of initial samples generated by stationary methods, and then new samples are sequentially added based on several criteria, to most effectively improve the surrogate model [23, 24]. Many criteria have been designed to balance the capabilities of exploration and exploitation [25,26].

The expected improvement function (EIF) [27] is a statistical criterion developed for Kriging models, which represents the expected improvement of the surrogate model when the new sample is placed at a certain location. Both exploration and exploitation can be achieved by maximizing the EIF, although the multiple local optima of the EIF can cause problems for the optimization process. The bumpiness function [28] is a similar, but computationally expensive, approach for RBF response surfaces. There is a more economical criterion that combines the power function and curvature for RBF response surfaces; this criterion can have a similar effect as the EIF has with Kriging [29]. The departure function [19] and prediction variance [30] have also been developed to evaluate the improvement to the surrogate model by adding a new sample. The concept of these methods is that the locations where samples are added should have the most significant impact on the surrogate model. Both the departure function and prediction variance have large computational costs because a large number of extra surrogates must be constructed.

**B. Adaptive sampling based on RBF response surfaces**

The OSS method is developed from a basic adaptive sampling algorithm based on RBF response surfaces. The RBF response surface is one of the most commonly used surrogate models with the capability of approximating high-dimensional nonlinear relationships. The OSS method utilizes a standard adaptive sampling algorithm, in which new samples are found through a multiobjective multiconstraint optimization. The optimization objectives, i.e., the proposed criteria for the OSS method, are introduced afterwards.

The RBF response surface reaches a relatively good balance between computational costs and predictive accuracy. The response surface is built on a sample set $\chi = \{(\mathbf{x}_i, \mathbf{y}_i)\}(i = 1,2,\ldots,M)$, where $\mathbf{x} \in \mathbb{D}^x \subseteq \mathbb{R}^{n_x}, \mathbf{y} \in \mathbb{D}^y \subseteq \mathbb{R}^{n_y}$, and $n_x$ and $n_y$ are the dimensions of the input space ($\mathbb{D}^x$) and output space ($\mathbb{D}^y$), respectively. The input vector is expressed as $\mathbf{x} = [x_1, \cdots, x_{n_x}]$. For cases in which $n_y = 1$, the RBF response surface gives the output estimation of $\mathbf{x}$ as the following equation:

$$s(\mathbf{x}) = \sum_{i=1}^{M} \beta_i \phi(\|\mathbf{x} - \mathbf{x}_i\|) + p(\mathbf{x}) \tag{1}$$

where $\|\cdot\|$ is the weighted and scaled Euclidean distance, i.e., $\|\mathbf{x} - \mathbf{x}_i\| = \sqrt{\frac{1}{\sum_{j=1}^{n_x} w_j^2} \sum_{j=1}^{n_x} w_j^2 (x_j - x_{i,j})^2}$. Function $\phi(\cdot)$ is the radial basis function, and $p(x)$ is an optional polynomial used to improve the RBF response surface quality. The detailed definitions can be found in reference [29]. When the optional polynomial is not applied, the weight vector $\boldsymbol{\beta} = [\beta_i]$ is determined by equation $\mathbf{y} = \mathbf{R}\boldsymbol{\beta}$, where $\mathbf{y} = [f_i]$ is the output vector, and $\mathbf{R} = [\phi_{ij}]$ ($\phi_{ij} = \phi(\|\mathbf{x}_i - \mathbf{x}_j\|)$, $i,j = 1,\cdots,M$) is the basis function matrix. Then, $s(x) = \mathbf{r}^T \boldsymbol{\beta}$, where $\mathbf{r} = [r_i], r_i = \phi(\|\mathbf{x} - \mathbf{x}_i\|)$. The coefficients $w_j$ and those in $\phi$ must be further determined to improve the model accuracy, and the root-square-mean-error (RSME) of leave-one-out cross-validation (LOOCV) is often used to optimize these coefficients.

The adaptive sampling algorithm of the OSS method is shown in Figure 2. After an initial sample set generated by stationary sampling methods such as iLHS, adaptive sampling is achieved by optimizations on surrogate models to generate new sample locations $\{\mathbf{x}_i\}$ ($i = 1, \ldots, N_{Add}$) in each iteration. The differential evolution (DE) algorithm [6] is used for optimizations in which the user-defined constraints of the inputs and outputs (i.e., $\mathbf{x} \in \mathbb{D}_x, \mathbf{y} \in \mathbb{D}_y$) are also applied. After the evaluation of these new samples, the sample set and surrogate models are updated. Substantially, the OSS is a multiobjective multiconstraint optimization process that is used to search for samples with different outputs and to improve the quality of surrogate models.

> Generate initial sample set $\{(\mathbf{x}_i, \mathbf{y}_i)\}(i = 1, \dots, N_{S,0})$;
> $N_S = N_{S,0}$;
> for $k = 1, \dots, (N_{S,max} - N_{S,0})/N_{Add}$ {
>   multiobjective optimization on RBF response surface {
>     objectives: criteria for adaptive sampling;
>     subject to: $\mathbf{x} \in \mathbb{D}_x, \mathbf{y} \in \mathbb{D}_y$;
>   }
>   Evaluation of new samples, i.e., $\{(\mathbf{x}_i, \mathbf{y}_i)\}$ $(i = 1, \dots, N_{Add})$;
>   Update sample set, $N_S = N_S + N_{Add}$;
>   Update RBF response surface;
> }
> end

**Figure 2 Adaptive sampling algorithm**

To generate multiple samples in each iteration, i.e., $N_{Add} > 1$ in Figure 2, usually two criteria are used as objectives for the optimization, and new samples are selected from the Pareto front of the final solutions. In addition, the initial population of the multiobjective optimization can be selected from the existing samples, which tends to have a better efficiency than randomly generating the initial population.

## C. Euclidean distance criteria

Several criteria for exploitation and exploration have been developed for adaptive samplings based on RBF response surfaces, e.g., separation and Laplacian functions [29, 31]. In contrast, the use of a much cheaper minimum Euclidean distance criterion, $d_{minX}$ (Eq. 2), can achieve comparable results with the separation function [32,33].

$$d_{minX} = \min_{1 \leq i \leq N_S} \|\mathbf{x} - \mathbf{x}_i\| \qquad (2)$$

Similarly, the space-filling property in the output space could be improved by maximizing a minimum Euclidean distance criterion in the output space:

$$d_{minY} = \min_{1 \leq i \leq N_S} \|\mathbf{y} - \mathbf{y}_i\| \qquad (3)$$

where $\|\mathbf{y} - \mathbf{y}_i\| = \sqrt{\frac{1}{n_y}\sum_{k=1}^{n_y}(y_k - y_{i,k})^2}$.

Another criterion $d_C$ can also be used for exploring the output space boundary, which is defined by Eq. 4 as the distance to the output space center $\mathbf{y}_C$. The output space center is estimated by the average output of all samples, i.e., $\mathbf{y}_C = \frac{1}{N_S}\sum_{i=1}^{N_S}\mathbf{y}_i$.

$$d_C = \|\mathbf{y} - \mathbf{y}_C\| \tag{4}$$

The Euclidean distance criteria are very intuitive, and they should be able to fill the output space and explore its boundaries. However, the sample distribution cannot be controlled, which may cause bias in statistical studies. Therefore, criteria that can effectively guide the sampling method to generate samples in desired places are needed. On the other hand, the quality of sampling is usually difficult to assess. In other words, a method to assess the extent to which a specified sample distribution is achieved is also needed. The space filling property is usually used to describe the quality of exploration or the quality of uniformly distributed samples. The crowding distance matrix [21] is a typical measure for describing the Euclidean distances between samples, which can represent the space filling property, to an extent. Similarly, the "potential energy" in Audze-Eglais design [22] also uses Euclidean distances to represent the space filling property. However, these criteria do not describe sample distributions, thus these Euclidean distance criteria are not sufficient for statistical studies.

**D. Target shooting criteria**

To specify the sample distribution in the output space, and improve the sampling efficiency, a target shooting criterion is proposed. The target shooting criterion utilizes targets in the output space to attract samples to be placed near them, so that the OSS method can more effectively obtain enough samples to achieve the specified sample distribution. Figure 3 shows the general idea of target shooting. First, $N_{target}$ targets ($\mathbf{y}_i^{target}$, i.e., the gray circles in Figure 3) are generated in the output space according to the specified sample distribution (e.g., a uniform distribution in Figure 3). Each target $\mathbf{y}_i^{target}$ attracts new samples to be placed in its neighborhood (i.e., the region between two gray dashed lines). The number of samples and the distances of these samples to $\mathbf{y}_i^{target}$ are described by its crowding level $\sigma_i$. The potential of each target ranges from 0 to 1, and it is designed to have a larger value when more samples and closer samples are generated in its neighborhood. Then, the summation of $\sigma_i$ shows the level of achieving the specified sample distribution. Therefore, by maximizing the $\sigma_i$ summation during the OSS, samples can be generated according to the specified distribution. In addition, due to the existence of an upper bound of the $\sigma_i$, the target will cease to attract new samples when its neighborhood is well filled, which can lead to an enhanced sampling efficiency.

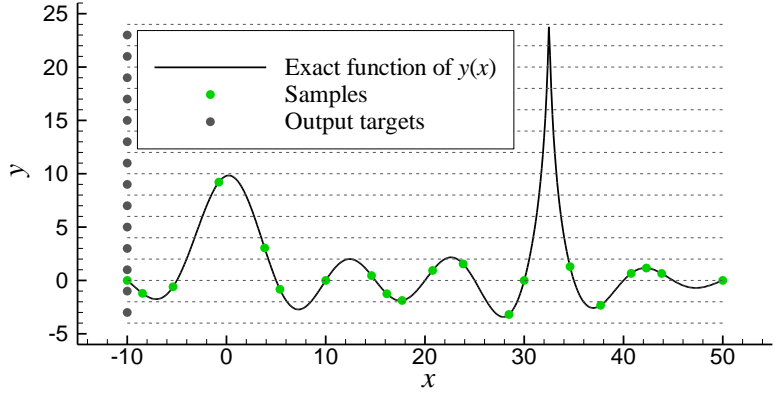

**Figure 3 Definition of the diversity measure**

To mathematically describe the target shooting process, a TSC based on targets and potential energy is proposed. For $M$ samples in the $n_y$-dimensional output space $\mathcal{F} = [0,1]^{n_y} \in \mathbb{R}^{n_y}$, a number of desired targets in the output space must be generated. A target set ($\mathbb{Y}_{target}$) with $N_{target}$ samples is generated according to the specified sample distribution using the existing sampling methods, e.g., the iLHS method is employed in the present paper to generate uniformly distributed targets. These output targets ($\mathbf{y}_i^{target} \in \mathbb{Y}_{target}, i = 1, \cdots, N_{target}$) are served as indicators for showing the crowding level ($\sigma$) of samples in each target's neighborhood. In addition, the preset number $N_{target}$ indicates how many different combinations of outputs are desired during the sampling.

The crowding level of indicator $\mathbf{y}_i^{target}$ is quantified as $\sigma_i$ when $M$ samples are presented:

$$\sigma_i = \max\left[1, \sum_{j=1}^{M} \epsilon(r_{ij})\right] \tag{5}$$

$$\epsilon(r) = (1 + c \cdot r)e^{-c \cdot r} \tag{6}$$

$$r_{ij} = \sqrt{\frac{1}{n_y}\sum_{k=1}^{n_y}(y_{i,k}^0 - y_{j,k}^0)^2} \tag{7}$$

where $c$ is a positive constant calculated by equation $\epsilon(\bar{r}_{target}) = 0.5$. The average scaled minimum distance between the samples in $\mathbb{Y}_{target}$ is defined as

$$\bar{r}_{target} = \frac{1}{N_{target}} \sum_{i=1}^{N_{target}} \left[\min_{j \neq i} r_{ij}\right] \tag{8}$$

Then, $\sigma_i \in [0,1]$ shows whether there are samples around the indicator $\mathbf{y}_i^{target}$ and how close the samples are to $\mathbf{y}_i^{target}$. Therefore, the TSC can be defined as Eq. 9, where $\Delta \sigma_i$ is the increase of $\sigma_i$ when a given sample is added, and ratio $\alpha_i$ will encourage the OSS to generate samples near the targets that have fewer samples in their neighborhood.

$$\Delta \sigma_{max} = \max_{1 \leq i \leq N_{target}} [\alpha_i \cdot \Delta \sigma_i] \tag{9}$$

$$\alpha_i = 1/\max(\sigma_i, 0.1) \tag{10}$$

Therefore, by maximizing $\Delta \sigma_{max}$, the targets with fewer nearby samples will more effectively attract new samples to be generated in the neighborhood. In addition, the targets whose neighborhoods are already well filled will cease to take effect.

**E. Diversity measure of samples**

The diversity of samples indicates whether the samples are sufficiently different from each other, which is also called the space-filling property. To quantify the diversity of the samples in the multidimensional output space, a diversity measure $\theta$ developed from the TSC is proposed, and it should meet the following three requirements [34]:

(1) $\theta \in [0,1]$, and $\theta$ should increase with increasing diversity;

(2) $\theta$ must not decrease when a new sample is added, and it should have a smaller increment when the new sample is more similar to existing samples;

(3) $\theta = 0$ means there are fewer than two samples in the sample set, and $\theta = 1$ represents a sufficient space-filling property.

To make the crowding level $\sigma_i$ meet the previous 1st and 2nd requirements, the potential energy $\epsilon$ must have an upper bound of 1 when $r$ approaches 0, and thus $\epsilon$ is described in Eq. 6. In addition, the diversity of sample set $\mathbb{Y} = \{\mathbf{y}_j\}$ $(j = 1, \cdots, M)$ can be roughly represented by

$$\tilde{\theta} = \frac{1}{N_{target}} \sum_{i=1}^{N_{target}} \sigma_i \tag{11}$$

To make the diversity $\theta$ meet the 3rd requirement, $\tilde{\theta}$ must be scaled by Eq. 12.

$$\theta = \frac{1}{\tilde{\theta}'_{target}} \frac{\tilde{\theta} - \tilde{\theta}_1}{1 - \tilde{\theta}_1} \tag{12}$$

where $\tilde{\theta}_1$ is the $\tilde{\theta}$ value of one random sample, and $\tilde{\theta}'_{target}$ is the $\tilde{\theta}$ value when $N_{target}$ samples generated by the same sampling method are in the output space. The $\tilde{\theta}'_{target}$ estimates the upper bound of $\tilde{\theta}$ when $N_{target}$ samples are generated, and the scaling of Eq. 12 can roughly make $\theta$ meet the 3rd requirement. Therefore, the diversity measure $\theta$ can be used as a quantitative measure to describe the extent to which the samples meet the specific distribution.

**F. Analytic tests**

With the proposed criteria, the OSS method is able to explore the output space, whereas stationary methods such as iLHS just fills the input space. The OSS method is tested on a two-dimensional function, and its results are compared with other stationary and adaptive sampling methods. The test function is described in Eq. 13, which has a circle output space for a square input space.

$$y_1 = r\cos[\pi(2x_2 - 1)]$$
$$y_2 = r\sin[\pi(2x_2 - 1)] \tag{13}$$
$$r = 1 - (|2x_1 - 1|)^{0.3}, \mathbf{x} \in [0,1]^2$$

Five cases are tested on this function, for which 400 samples are generated and plotted in Figure 4. The iLHS method is the baseline sampling method, and the adaptive sampling method using the criteria in [32] represents the conventional adaptive sampling methods. The OSS-1 case employs an output space sampling with the Euclidean criteria $d_{minX}$ and $d_{minY}$, and the OSS-2 case uses the criteria $d_C$ and $\Delta\sigma_{max}$. The fifth case is an optimization minimizing output $y_1$ and $y_2$. The other four cases all start from the same initial sample set, which contains 100 samples generated by using the iLHS method. For adaptive samplings, i.e., cases 2, 3, and 4, four samples are generated in each iteration. The optimization uses a multiobjective DE algorithm with a population size of 16. The diversity histories of the four samplings and the theoretical diversity upper bound are shown in Figure 5 and Table 1.

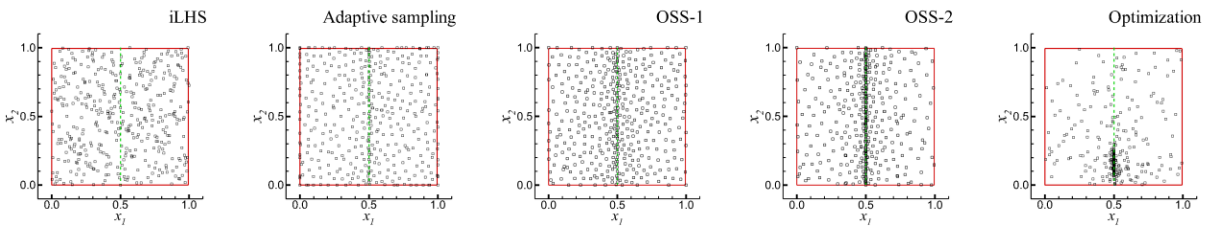

**(a) Sample distributions in the input space**

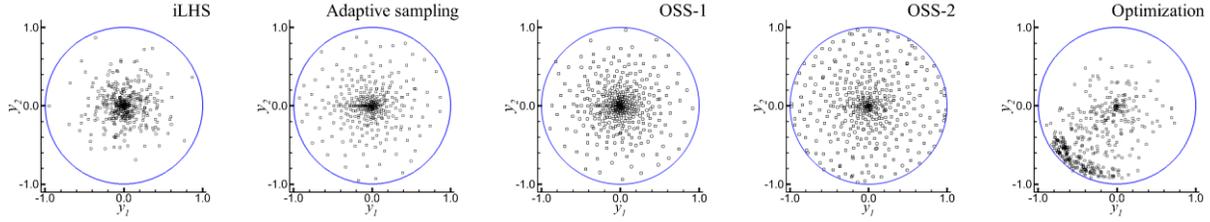

**(b) Sample distributions in the output space**

**Figure 4 Sample distributions of the five cases**

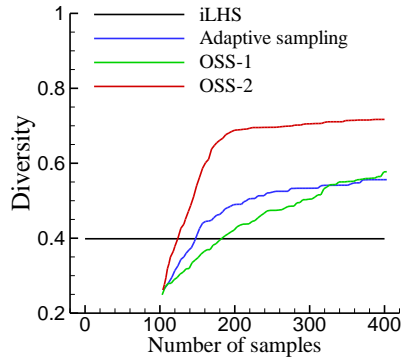

**Figure 5 Diversity history of the four sampling processes**

Table 1 Output diversities of the three methods

| Method | iLHS | Adaptive sampling | OSS-1 | OSS-2 | Theoretical |
|---|---|---|---|---|---|
| Diversity | 0.40 | 0.56 | 0.58 | 0.71 | 0.79 |

The results show that the iLHS method generates evenly spread samples in the input space; thus, the samples only cover a part of the output space. The adaptive sampling method can explore the output space to a certain extent, but the boundaries are not explored. The optimization can also explore the boundaries by forming a Pareto front, which covers approximately a quarter of the entire output space boundary. However, many of the samples generated by the optimization are very similar, and the output space remains not fully explored. In contrast, the OSS method can not only fill the output space but can also explore the boundaries by using different criteria. The OSS-1 case uses the Euclidean criteria $d_{minX}$ and $d_{minY}$ to explore the input and output spaces, which has a similar effect as that of the conventional adaptive sampling methods. Whereas the OSS-2 case uses the $d_C$ criterion to explore the output boundaries and the $\Delta\sigma_{max}$ criterion to fill the output space, and it achieves the best sample diversity.

## III. Statistical study of the supercritical airfoils

Korn's equation identifies the connections between the drag divergence Mach number $M_{\infty,DD}$, lift coefficient $C_L$ and airfoil maximum thickness $(t/c)_{max}$ at transonic speeds, which is described as $M_{\infty,DD} + 0.1C_L + (t/c)_{max} = \kappa$. $\kappa$ represents the airfoil technology level of different airfoils. For NASA supercritical airfoils, $\kappa$ is approximately 0.95, and it shows a good estimation at lift coefficients of approximately 0.4 and 0.7 [4].

Korn's equation shows a simple linear correlation between $M_{\infty,DD}$, $C_L$ and $(t/c)_{max}$; however, the parameter $\kappa$ raises a question regarding the influence of other airfoil features that must be answered. Because the flight performances of supercritical airfoils at transonic speeds are significantly affected by pressure distributions, their influences to technology level $\kappa$ must be studied further.

This section employs the OSS method to generate supercritical airfoils of the same $(t/c)_{max}$, $C_L$, same shock wave strength and similar lower surfaces, to study the influence of shock wave location and suction peak. Their effects on the drag creep characteristic are also studied.

### A. Airfoil parameterization and CFD methods

The class shape function transformation (CST) method is selected to generate the airfoil geometry. The CST method combines class functions and shape functions to describe an arbitrary geometry and can guarantee airfoil smoothness with comparatively fewer design variables [35]. A sixth-order Bernstein polynomial is used as the shape function; i.e., seven CST parameters are used to describe upper and lower surfaces.

A C-grid is employed for Reynolds Average Navier–Stokes (RANS) analysis. The CFD solver is a well-known open source solver named CFL3D [36]. In this paper, the MUSCL scheme, Roe's scheme, the lower-upper symmetric Gauss-Seidel method, and the shear stress transport (SST) model are selected for reconstruction, spatial discretization, time advancing, and turbulence modeling, respectively, in RANS.

Figure 6 shows the experimental pressure coefficient ($C_p$) distributions of an RAE2822 airfoil [37] compared with the CFD results of different grid sizes. Two flight conditions with weak and stronger shock waves are used for validation. Figure 6 (a) shows that the free-stream Mach number ($M_\infty$) is 0.725, the Reynolds number ($Re$) based on the unit chord length is $6.5 \times 10^6$, and the angle of attack ($AoA$) is 2.55 degrees. Figure 6 (b) shows that $M_\infty$ is 0.73, $Re$ is $6.5 \times 10^6$, and $AoA$ is 3.19 degrees. The three grids, with sizes of 20,000, 40,000, and 80,000, have 201, 301, and 401 grid points on the airfoil surface, respectively. The $\Delta y+$ of the first grid layer is always kept less than 1. All grids can achieve a similar $C_p$ resolution for both cases.

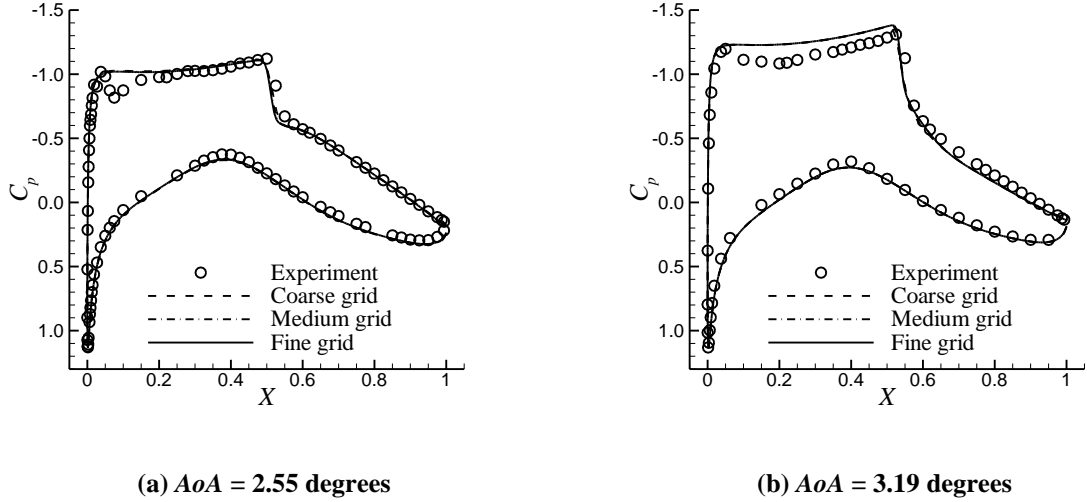

(a) *AoA* = 2.55 degrees  (b) *AoA* = 3.19 degrees

**Figure 6 Grid convergence study of $C_p$**

**B. Key pressure distribution features**

The performances of an airfoil are comprehensively associated with its pressure coefficient distributions, which can also be described in wall Mach number distributions. The wall Mach number ($M_w$) is the Mach number calculated based on an isentropic relationship with the pressure coefficient on the airfoil surface and free-stream Mach number $M_\infty$ [38]. The following key parameters are illustrated in Figure 7.

(1) Wall Mach number of the suction peak, $M_{wL}$: the wall Mach number of the suction peak on the upper surface, which is defined as the largest wall Mach number within the 15% chord length near the leading edge.

(2) Location of the shock wave, $X_1$: initially, the shock wave is roughly located by the largest $-dM_w/dX$ location $\tilde{X}$ on the upper surface. Then, the largest $-d^2M_w/dX^2$ location is found in front of $\tilde{X}$ as $X_1$. For practical use, the local wall Mach number of $X_1$ should be greater than 1, and the largest $-dM_w/dX$ must exceed a specific predefined criterion. Otherwise, it is not a shock wave at this candidate $X_1$ location. The criterion is related to the number of grid points on the upper surface and the free-stream Mach number.

(3) Wall Mach number in front of the shock wave, $M_{w1}$: the $M_w$ of location $X_1$.

(4) Highest wall Mach number on the lower surface, $M_{w,lower}$.

(5) Smoothness of the suction plateau, $Err$: smoothness function of the suction plateau is defined as the area of the shaded region between the suction peak and shock wave. It is used to avoid airfoils on which the suction plateau has a severe pit or fluctuation. The smoothness function can be defined for both the wall Mach number distribution and the pressure coefficient distribution ($C_p$). The smoothness of the pressure coefficient distribution is used in the present paper.

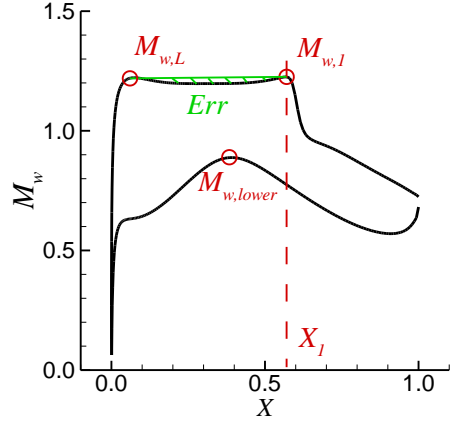

**Figure 7 Definitions of pressure distribution features**

### C. Airfoil sampling of shock wave locations

The OSS method is used to generate airfoils with the given constraints. The airfoils have the same $(t/c)_{max}$ of 0.095, and the adaptive sampling procedure is conducted under cruise conditions, i.e., free-stream Mach number $M_\infty$=0.76 and Reynolds number $Re = 5 \times 10^6$ based on a chord length of 1. The flight angle of attack is adjusted by keeping a fixed lift coefficient of $C_L$=0.70. Several of the constraints described in Eq. 14 are applied for better generalization ability in practical applications, and the $M_{w,lower}$ is restricted to values less than 0.9 to avoid the supersonic region on the lower surface when the free stream Mach number is increased.

$$\text{s.t. single shock wave, } Err < 0.02, M_{w,lower} < 0.9 \qquad (14)$$

To study the influence of shock wave location, the shock wave strength is fixed for sampling, i.e., the wall Mach number in front of the shock wave ($M_{w,1}$) is roughly the same for these airfoils. In this section, airfoils that have $M_{w,1} \approx 1.11$ (i.e., $M_{w,1} \in [1.11,1.12]$ ), $M_{w,1} \approx 1.14$ (i.e., $M_{w,1} \in [1.14,1.15]$ ) and $M_{w,1} \approx 1.16$ (i.e., $M_{w,1} \in [1.16,1.17]$) are generated and studied.

The initial sample set contains 400 airfoils selected from the cruise drag reduction optimization described in Eq. 15. The optimization has 32 individuals in a population, and 100 generations are carried out for complete convergence. After the optimization, there are 2000 airfoils meet the constraints in Eq. 14. However, to ensure that the airfoils are not too similar for the initial set of OSS sampling, 400 valid airfoils from the entire optimization process are selected instead of using airfoils from the few most recent generations.

$$\min C_d(M_\infty = 0.76, C_L=0.70)$$
$$\text{s.t. single shock wave, } Err < 0.02, M_{w,lower} < 0.9, C_d < 0.014 \tag{15}$$

Starting from the 400 airfoils from the optimization, four samples are generated in each iteration during the OSS sampling process. The $d_{minX} + d_C$ and $\Delta\sigma_{max}$ criteria are used as the two objectives to pursue the diversity of the shock wave location ($X_1 \in [0.3,0.7]$), and the surrogate model quality is also improved during this process. The $N_{target}$ for the target shooting criterion $\Delta\sigma_{max}$ is set to 40. Figure 8 shows the diversity and valid sample number histories during the sampling processes, as well as the increased number of valid samples. This figure also indicates that the diversity has mostly converged after 100 generations of sampling.

After neglecting similar samples to maintain the balance of sample numbers in different intervals, Figure 9 shows the distribution of valid samples in each sampling, where the scattered points represent the exact number of samples in each interval, and the dashed curves are the estimated distribution of the samples. The wall Mach number distributions of these airfoils are plotted in Figure 10 as gray lines, where the gray shade further confirms that the samples effectively cover the $X_1$ ranges of [0.39,0.70], [0.34,0.68] and [0.34,0.68] in the $M_{w,1} \approx 1.11$, 1.14 and 1.16 cases, respectively. Therefore, the results show that the OSS method can effectively generate samples to explore the pressure distribution space under constraints.

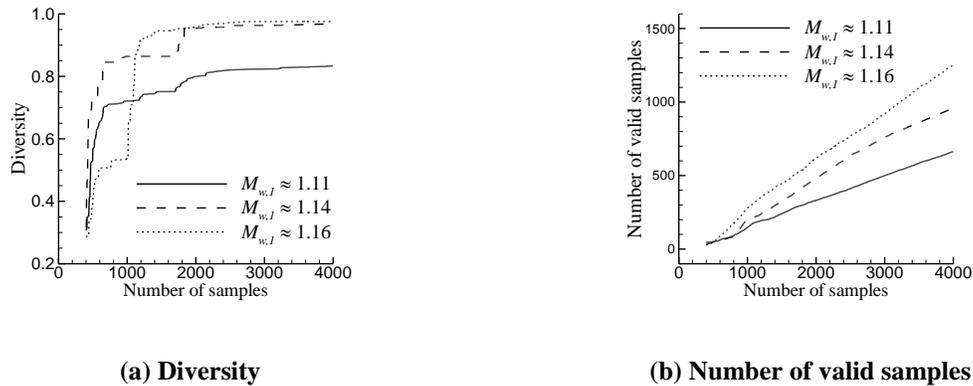

(a) Diversity          (b) Number of valid samples

**Figure 8 History of the adaptive sampling processes**

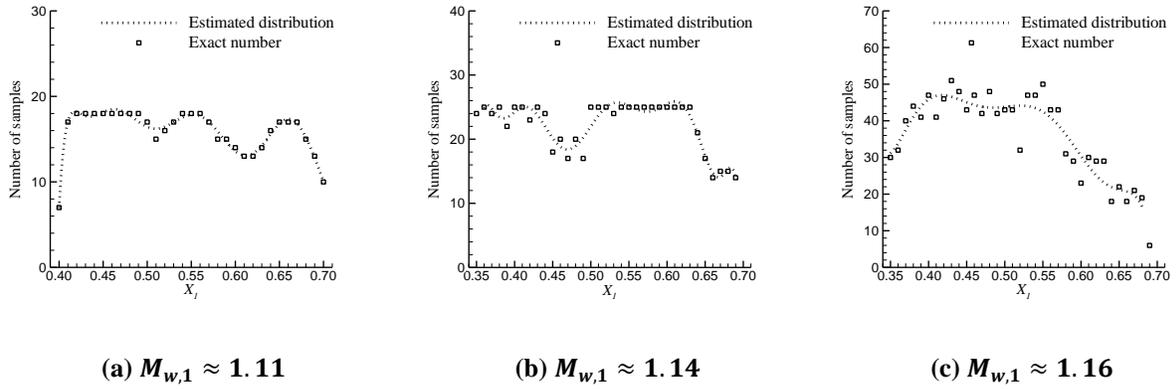

(a) $M_{w,1} \approx 1.11$      (b) $M_{w,1} \approx 1.14$      (c) $M_{w,1} \approx 1.16$

**Figure 9 Sample distribution of the adaptive sampling results**

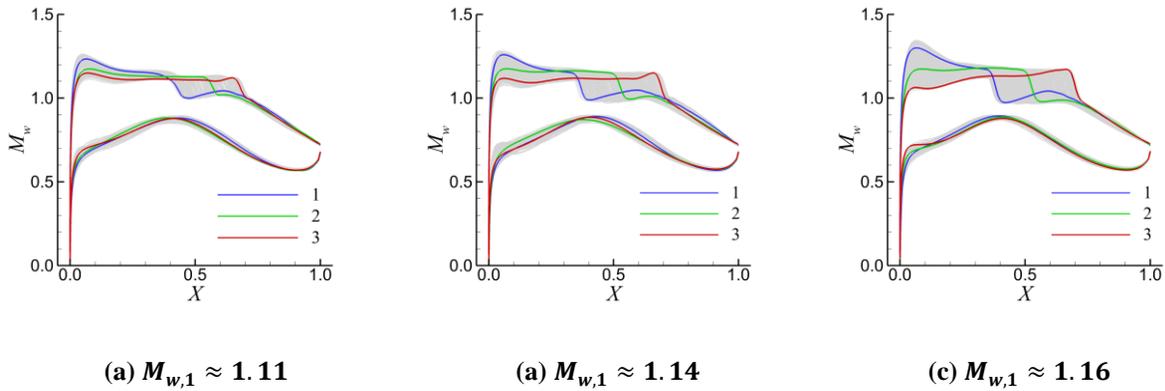

(a) $M_{w,1} \approx 1.11$      (a) $M_{w,1} \approx 1.14$      (c) $M_{w,1} \approx 1.16$

**Figure 10 Valid airfoil samples and typical samples**

### D. Statistical analysis of airfoil performances

The airfoils are evaluated at several different free stream Mach numbers when the lift coefficient and Reynolds numbers are held the same as in the cruise condition. The off-design free stream Mach numbers are 0.700, 0.720, 0.740, 0.750, 0.770, 0.775, 0.780, 0.785, and 0.790. Three typical airfoils are selected in each case, which are shown as airfoils 1, 2, and 3 in Figure 10, and their $C_d - M_\infty$ curves are plotted in Figure 11. The figure shows that the shock wave location or suction peak has significant influence on the drag divergence and drag creep. The influence can vary based on shock wave strength, and the influence grows stronger when the shock wave is more upstream. To further reveal the influence of these flow features, the drag divergence Mach number and drag increase describing drag creep are studied with different values of flow features.

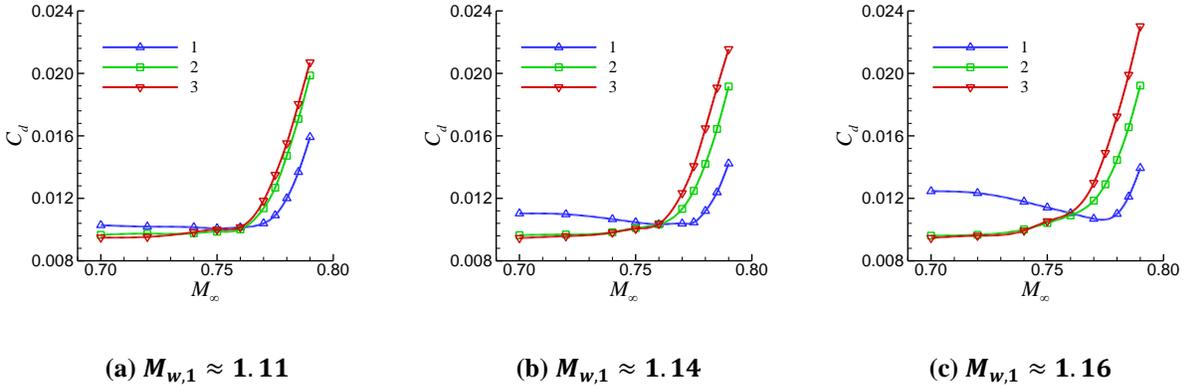

(a) $M_{w,1} \approx 1.11$  (b) $M_{w,1} \approx 1.14$  (c) $M_{w,1} \approx 1.16$

**Figure 11 Drag – Mach number curves of typical airfoils**

Drag divergence Mach number can be defined as the point on a curve of $C_d - M_\infty$ at constant $C_L$ at which the slope becomes 0.1 for aircrafts or wings [39]. Considering a supercritical wing with a swept angle of 30 degrees, when using a simple sweep theory [39], the critical slope $k$ becomes $(\partial C_d/\partial M_\infty)_{2D} = (\partial C_d/\partial M_\infty)_{3D}/\cos^4 30° = 0.18$ for 2-D airfoils. According to Figure 12, the drag divergence Mach numbers determined by $k$ are equal to 0.1, 0.2, and 0.3 in both cases. This figure also shows that the drag divergence Mach number $M_{\infty,DD}$ decreases when the shock wave is further downstream, while indicating that the linear correlation is mostly maintained for different values of $k$.

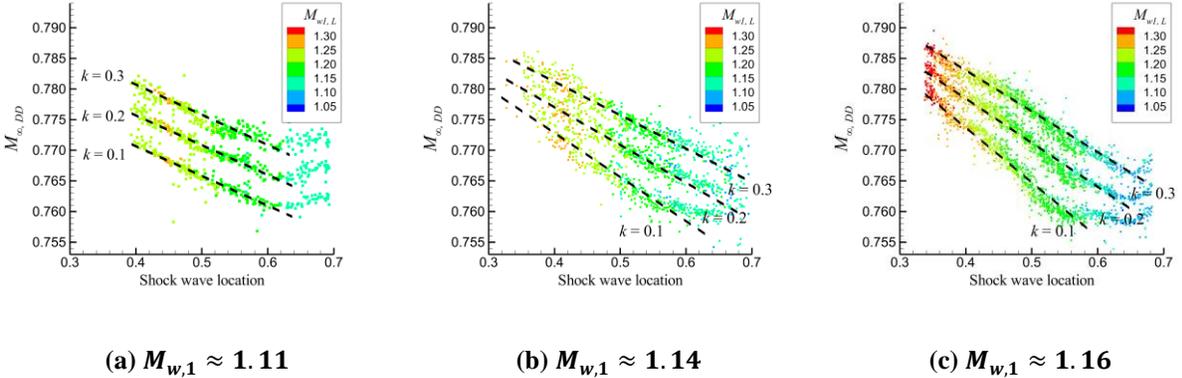

(a) $M_{w,1} \approx 1.11$  (b) $M_{w,1} \approx 1.14$  (c) $M_{w,1} \approx 1.16$

**Figure 12 Influence of shock wave location on $M_{\infty,DD}$**

Meanwhile, since $M_{w,L}$ shows a negative correlation with $X_1$, as shown in Figure 13, the $M_{\infty,DD}$ also increases when the suction peak is greater. Figure 14 shows that $M_{\infty,DD}$ has a stronger correlation with shock wave location $X_1$ than the $M_{w,L}$ of the suction peak, for which the R-squared ($R^2$) values of the two linear regressions are 0.932 and 0.795, respectively. Because all these airfoils share the same lift coefficient and maximum relative thickness, the difference in $M_{\infty,DD}$ indicates the influence of the pressure distribution features on the technology level factor $\kappa$. Therefore, the technology level factor can be described by Eq. 16.

$$\kappa = 0.97 - 0.065X_1 \ (R^2 = 0.0932) \tag{16}$$

In other words, Korn's equation can be further described as Eq. 17.

$$M_{\infty,DD} + 0.1C_L + (t/c)_{max} + 0.065X_1 = 0.97 \tag{17}$$

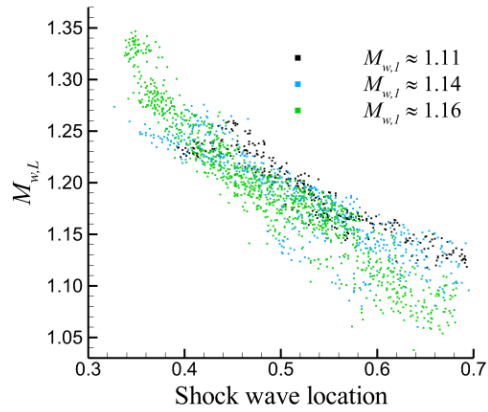

**Figure 13 Correlation between shock wave location $X_1$ and $M_{w,L}$**

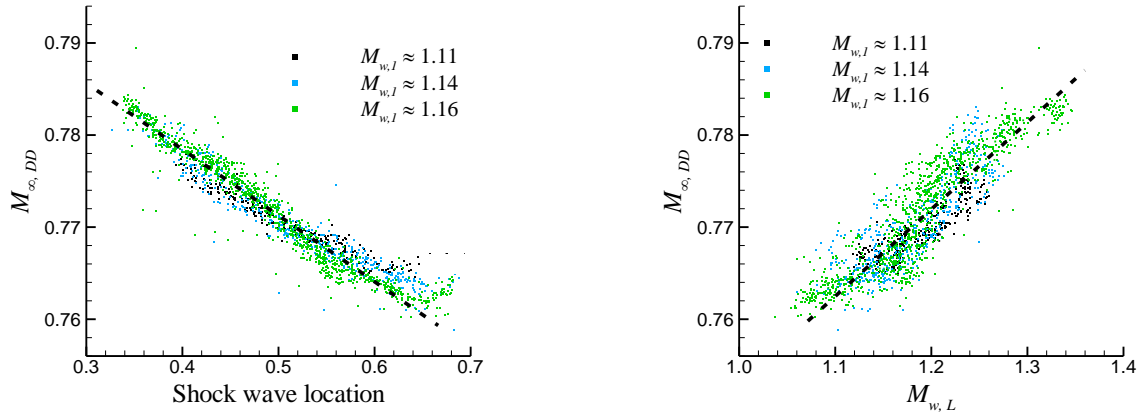

(a) shock wave location ($R^2$=0.932)     (b) wall Mach number of suction peaks ($R^2$=0.795)

**Figure 14 Influence of flow features on $M_{\infty,DD}$ ($k$ = 0.2)**

In contrast, the drag creep characteristic of the airfoils does not have a linear relationship with $X_1$ or $M_{w,L}$, although the correlation is comparatively stronger with $M_{w,L}$ than $X_1$. Figure 15 shows the influence of $X_1$ and $M_{w,L}$ on the drag creep characteristic, which is represented by the drag coefficient increment between $M_\infty$ = 0.76 and 0.70, i.e., $\Delta C_{d,creep} = C_{d,M_\infty=0.76} - C_{d,M_\infty=0.70}$. Although a simple regression is not easy to achieve, nonetheless this figure generally indicates that the drag coefficient rapidly grows when the $M_{w,L}$ values of the airfoils are too high. Therefore, to avoid bad drag creep characteristics, an airfoil should have its $M_{w,L}$ lower than 1.25, as shown in Figure 15 (b).

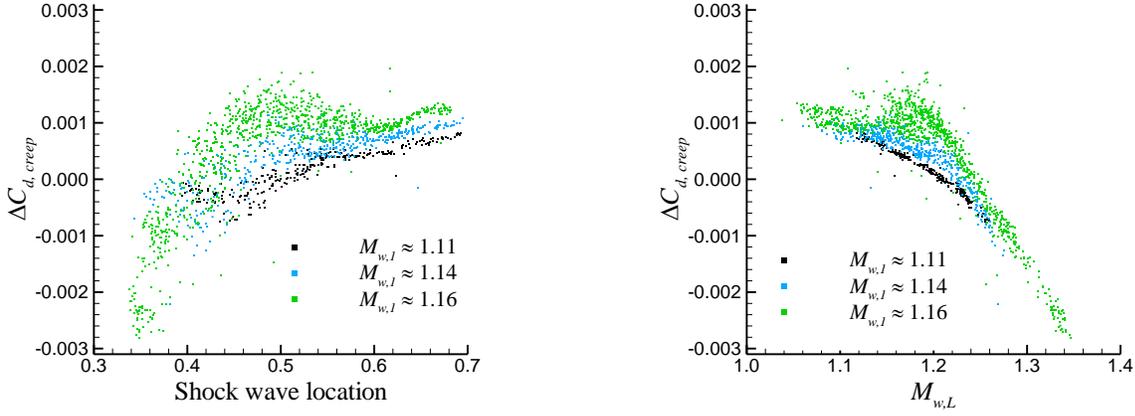

(a) shock wave location  (b) wall Mach number of suction peaks

**Figure 15 Influence of flow features on the drag creep characteristic**

## IV. Conclusions

The output-space sampling method is important for statistical aerodynamics research on the relationship between flow features and performance. As part of this method, here, a quantitative assessment of the sample diversity in the output space was proposed, its boundary exploitation capability was enhanced, and the space-filling objective was also effectively realized in the output space. The sampling method was proven to be more efficient for guaranteeing the diversity of flow features than the conventional sampling methods.

The output space sampling method was then used to generate airfoils with various pressure distribution features. This method was able to efficiently obtain unbiased and consistent sample distributions under different constraints. This property further guaranteed the accuracy and generality of the statistical research.

Supercritical airfoils generated by the OSS method were statistically analyzed. The results showed that the drag divergence Mach number increased when the shock wave was more upstream and that the suction peak became higher as well. Korn's equation was improved to $M_{\infty,DD} + 0.1C_L + (t/c)_{max} + 0.065X_1 = 0.97$. The drag creep characteristic may deteriorate if the suction peak is greater than $M_{w,L} = 1.25$. These relations and constraints can guide the selection of the shock wave location during the process of aerodynamic optimization and design.

## Acknowledgments

This research was supported by the National Natural Science Foundation of China under grant Nos. 91852108 and 11872230 and the Tsinghua University Initiative Scientific Research Program under grant No. 2015Z22003.